\documentclass[prl,twocolumn,showpacs,preprintnumbers,amsmath,amssymb,floats]{revtex4-1}
\pdfoutput=1
\usepackage{graphicx}
\usepackage{amsmath}
\usepackage{epstopdf}
\usepackage{amsbsy}
\usepackage{color}
\usepackage{dcolumn}
\usepackage{bm}

\newcommand\rr{{\mathbf r}}
\newcommand\R{{\mathbf R}}
\usepackage{soul}

\begin{document}

\title{Visualization of atomic-scale phenomena in superconductors: application to FeSe }
\author{Peayush Choubey$^1$, T. Berlijn$^{1,2}$, A. Kreisel$^1$, C. Cao$^3$ and  P. J. Hirschfeld$^1$}
\affiliation{
$^1$Department of Physics, University of Florida, Gainesville, Florida 32611, USA\\
$^2$Center for Nanophase Materials Sciences and Computer Science and Mathematics Division, Oak Ridge National Laboratory, Oak Ridge, Tennessee 37831, USA
\\
$^3$Condensed Matter Physics Group, Department of Physics,
Hangzhou Normal University, Hangzhou 310036, China}

\date{\today}

\begin{abstract}
We propose a simple method of calculating inhomogeneous, atomic-scale phenomena in superconductors which makes use of the  wave function information traditionally discarded in the construction of tight-binding models used in the Bogoliubov-de Gennes equations.  The method uses symmetry-based first principles Wannier functions to visualize the effects of superconducting pairing on the distribution of electronic states over atoms within a crystal unit cell.    Local symmetries lower than the global lattice symmetry can thus be exhibited as well, rendering theoretical comparisons with scanning tunneling spectroscopy data much more useful.    As a simple example, we discuss the geometric dimer states observed near defects in superconducting FeSe.
\end{abstract}


\maketitle

In the past two decades, two developments have spurred new interest in atomic-scale effects in superconductors.  The first is the discovery of systems, like the high-$T_c$ cuprates, heavy fermion, and Fe-based superconductors, with very short coherence lengths $\xi$, which in the cuprates approach the unit cell size.   The second is the ability to image atomic- and even sub-atomic scale features in scanning tunneling spectroscopy (STS). 
Comparison of such images with theory has sometimes been frustrating, since STS images often contain much more detail than that predicted by current calculations.

 The success of the semiclassical theory of superconductivity\cite{Eilenberger,RainerSerene}, which integrates out rapid oscillations of the pair wave function at the atomic-scale, has led to the widespread notion that atomic scale effects are irrelevant for superconductivity.
However, observation of the electronic structure very near a defect or surface reflects directly the symmetry and structure of the superconducting order parameter, and can furthermore provide information on   gap inhomogeneity, vortex structure,
 competing phases, and even the atomic-scale variation of the pairing interaction.  Yet in the currently available method used to describe inhomogeneous phenomena, the Bogoliubov-de Gennes equations,  atomic wave function information is typically integrated out by adopting a tight-binding
model for the normal state electronic structure.

   There is now a growing list of problems which seem to require atomic scale ``details'' beyond  the BdG description usually employed.    For example, STS images of underdoped cuprates show the existence of unusual incommensurate electronic modulations, proposed to represent real-space images of the pseudogap, which have alternating high and low intensity on O-centered bonds between neighboring Cu's\cite{Kapitulnik1,Davis_checkerboard1,Davis_checkerboard2}; such effects cannot be captured by conventional models which include Cu states only.    
In general there has been a great recent effort to map out the intra-unit cell electronic structure of unconventional superconductors \cite{Lee,Mesaros,Hamidian,Fujita,Lawler}.
As a second example, there is still no consensus on the origin of the STS pattern around a Zn impurity in cuprates, which displays a central intensity maximum on the Zn site, unlike simple models of the response of a d-wave superconductor to a strong potential\cite{Zn}.  A popular explanation relies on ``filter" effects which describe a nontrivial tunneling process from the surface to the CuO plane\cite{Ting,Martin}, but while there is some support for these ideas from first principles calculations in the normal state\cite{abinitioZn}, there has been until now no way to include both superconductivity and the various atomic wave functions in the barrier layers responsible for the filter\cite{Markiewicz}.    Finally,  in Fe-based superconductors a variety of defect states observed by STS in FeSe, LiFeAs, NaFeAs and other materials\cite{song11,grothe12,hanaguri12,zhou11,song12,pasupathy13} break the square lattice symmetry locally and therefore cannot be
described by Fe-only lattice BdG calculations. These include so-called ``geometrical dimer'' structures aligned with the As or Se states at the surface.  
Note that in all of the above cases measurements sensitive to the LDOS at the surface are being used to analyze physics taking place several \AA\, beneath the surface, without a full understanding of the states involved in the tunneling process. 
Given that the STM tip almost always resides far away from the electronically active layers, these types of problems are to be expected for many superconducting materials.

    In this work, we present
a simple technique which restores the intra-unit cell information lost in the construction of the Bogoliubov-de Gennes Hamiltonian, including the degrees of freedom corresponding to other atoms in the unit cell, as well as neighboring cells.   In weak-to-intermediate coupling systems,  the method  dramatically improves the ability of theory to quantitatively describe intra-unit cell  problems, as we illustrate by presenting the solution to the FeSe problem described above.
In strongly coupled systems like cuprates,  it should still be valid for symmetry-related questions.

We first introduce the elementary expression for the local continuum Green's function in terms of the BdG lattice Green's function which is needed for the calculation of the  local density of states (LDOS) relevant for the STS experiments.   We then show that the \textit{nonlocal}  terms  also contribute and can be important even for the \textit{local} continuum LDOS.
We present a tight-binding model for FeSe derived from Density Functional Theory (DFT) and 
corresponding Wannier basis.   Within a 10-Fe-orbital BdG calculation, we calculate the response of the superconductor to a single pointlike impurity potential on an Fe site, and compare to the continuum LDOS calculated using the Wannier states.  The geometric dimer states are shown to emerge naturally from this description.  Finally, we discuss the approximations that we have made, as well as how the method can be improved and extended in the future.

\begin{figure}
\includegraphics[width=\columnwidth]{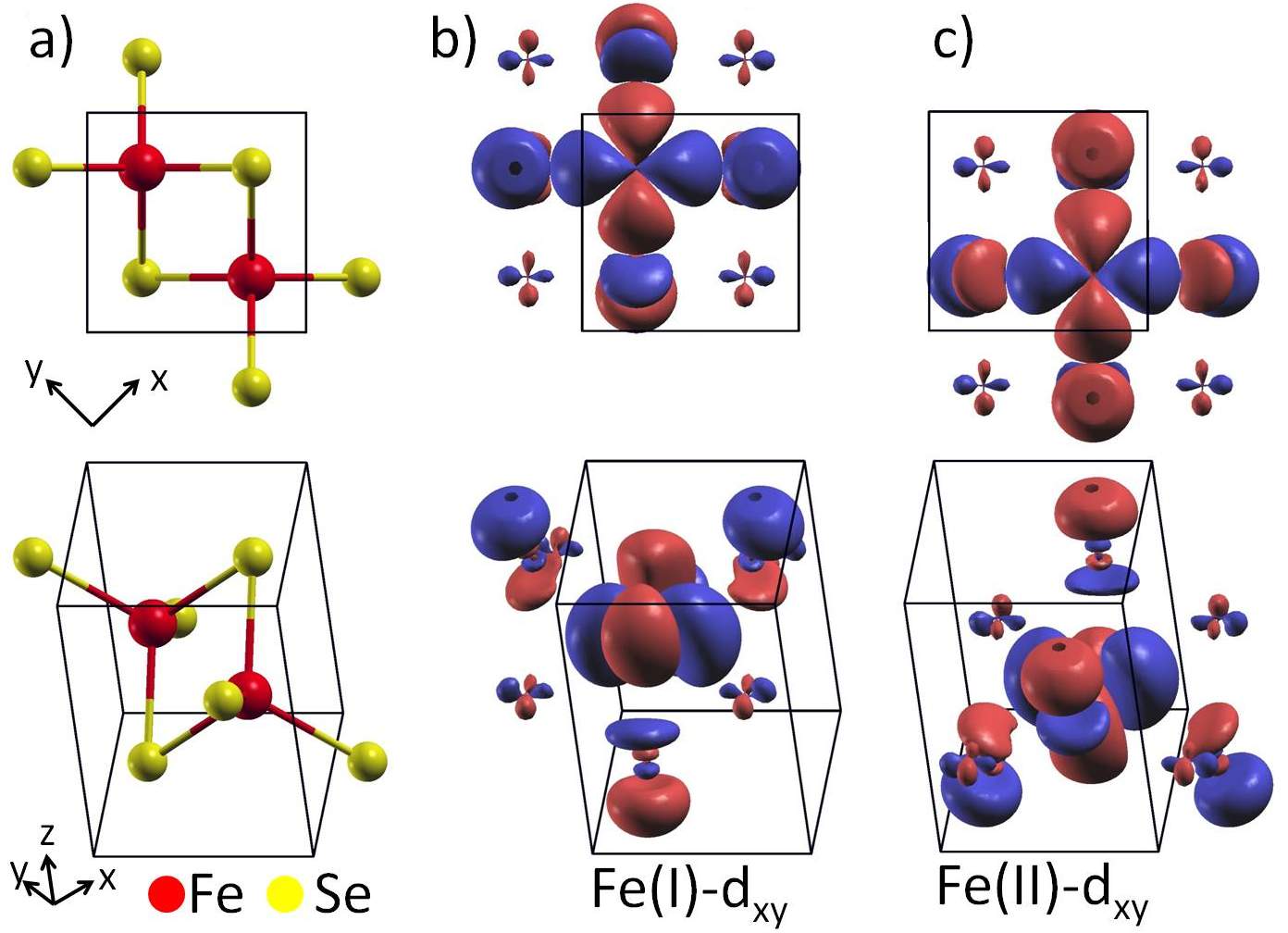}
\caption{(color online)  Isosurface plots of Fe-$d$ Wannier orbitals in FeSe at $0.03\,\text{bohr}^{-3/2}$;  red and blue indicate phase of the wave function.  a) (top) Top view of unit cell containing two Fe atoms (red) and two Se atoms (yellow), plus four Se exterior to unit cell; (bottom) side
view of same cell.    b)  (top) Top view of $d_{xy}$ Wannier orbital on Fe(I) site; (bottom) side view of same orbital. c) Same as b) but on Fe(II) site. Images were produced
with XCRYSDEN\cite{Kokalj}.}
\label{fig:wannier}
\end{figure}

 \textit{Method}.  The tight binding model, while often treated as phenomenological, is in principle derived from Wannier orbitals based on first principles calculations.  In order to preserve the desired mutual orthogonality of these orbitals, they are constructed from atomic wave functions on atoms in more than one unit cell, although they are exponentially localized.  In a crystal with more than one type of atom per unit cell, Wannier orbitals may represent linear combinations of wave functions from more than one atomic species.  Here we first construct from the DFT wave functions a  basis $w_{\R\mu}(\rr)$ using a projected Wannier method which preserves the local symmetry of the atomic states\cite{Ku_Wannier,Anisimov}.
   Here $\mu$ is an orbital index, $\R$ labels the unit cell, and $\rr$ describes the continuum position.
   In Fig. \ref{fig:wannier}, we show examples of the  Wannier  Fe-$d$ orbitals on the Fe(I) and Fe(II) sites derived for the homogeneous FeSe system,
  using the Fe-$d$ bands within the energy range $[-2.5,3]\,\text{eV}$ obtained from DFT  using Wien2K\cite{Blaha}.
  It is clear that, while these functions have features easily associated with the form of the corresponding atomic orbitals,
  they are considerably more complex, and  involve significant contributions from the Se states integrated out in the downfolding, with clear differences between
  Fe(I) and Fe(II).   These Wannier functions are  used to derive the tight-binding model containing Fe site energies and hoppings in the usual way
 to yield the Hamiltonian  $H_0=\sum_{\mathbf{RR'},\mu\nu,\sigma}t_{\mathbf{{\mathbf{RR}'}}}^{\mu\nu}c_{\mathbf{R}\mu\sigma}^{\dagger}c_{\mathbf{R'}\nu\sigma}-\mu_0\sum_{\mathbf{R}\mu\sigma}n_{\mathbf{R}\mu\sigma}$,
 where
$t_{\mathbf{RR'}}^{\mu\nu}$ are hopping elements between orbitals
  $\mu$ and $\nu$ in  unit cells $\mathbf R$ and $\mathbf R'$.
The full Hamiltonian for a superconductor in the presence of an impurity is therefore
 \begin{equation}
 \label{eq:H}
 H=H_{0}+H_\text{BCS}+H_\text{imp},
\end{equation}
where
 \begin{equation}
 H_\text{BCS}=-\sum_{\mathbf{R}, \mathbf{R'},\mu\nu}\Delta_{\mathbf{RR'}}^{\mu\nu}c_{\mathbf{R}\mu\uparrow}^{\dagger}c_{\mathbf{R'}\nu\downarrow}^{\dagger}+H.c.,
\end{equation}
with the superconducting order parameter  \mbox{$\Delta_{\mathbf{RR'}}^{\mu\nu}=\Gamma^{\mu\nu}_{\mathbf{RR'}}\langle c_{\mathbf{R'}\nu\downarrow}c_{\mathbf{R}\mu\uparrow}\rangle$},
and
\begin{equation}
   H_\text{imp}=\sum_{\mu^*\sigma}  V_\text{imp} c^\dagger_{\mathbf{R}^* \mu^*\sigma}c_{\mathbf{R^*} \mu^*\sigma},
\end{equation}
where   $\mathbf{R^*}$ is the impurity unit cell,  $\mu^*$ in this term runs only over the 5$d$ orbitals associated with the impurity Fe site, and for simplicity we have taken the impurity potential $V_\text{imp}$ to be nonmagnetic and  proportional to the identity  in the orbital basis.   Here  $\Gamma_{\mathbf{RR'}}^{\mu\nu}$ is the pairing interaction in orbital and real space.
The inhomogeneous mean field BdG equations for this Hamiltonian may now be solved by diagonalization with the auxiliary self-consistency equation for the gap $\Delta_{{\mathbf{RR'}}}^{\mu\nu}$ to obtain the BdG eigenvalues $E_{n\sigma}$ and eigenvectors $u_{n\sigma}$ and $v_{n\sigma}$.
From these we can construct the usual  retarded lattice Green's function
 \begin{equation}
G^{\mu\nu}_\sigma(\R,\R';\omega)\!=\! \sum_n\! \left(\frac
{u^{n\sigma}_{\R\mu} u^{n\sigma *}_{\R'\nu}}{ \omega-E_{n\sigma} +i0^+ }
+
\frac{v^{n-\sigma}_{\R\mu} v^{n-\sigma *}_{\R'\nu}}{ \omega+E_{n-\sigma} +i0^+  }
\right).
\end{equation}

 Under a wide set of conditions\cite{Hamman}, an STS experiment at bias $V=\omega/e$ measures the local density of states $\rho(\rr,\omega)\equiv -\frac 1 \pi\mathop\text{Im}G(\rr,\rr;\omega )$, related to the  retarded continuum  local Green's function $G(\rr,\rr';\omega )$.   

To relate the lattice and continuum Green's functions, one simply performs a basis transformation from the lattice operators $c_{\R\mu\sigma}$ to the continuum operators $\psi_\sigma(\rr)=\sum_{\R\mu} c_{\R\mu\sigma}w_{\R\mu}(\rr)$ where the Wannier functions $w_{\R\mu}(\rr)$ are the matrix elements,
\begin{equation}
G(\rr,\rr';\omega) =\sum_{\R,\R',\mu,\nu} \hskip -0.2cm G^{\mu\nu}(\R,\R';\omega) w_{\R\mu}(\rr) w^*_{\R'\nu}(\rr').
\label{eq:continuumG}
\end{equation}
Note that even  the local continuum Green's function $G(\rr,\rr;\omega)$  includes nonlocal and orbitally nondiagonal lattice Green's function
terms $G^{\mu\ne\nu}(\R\ne \R'; \omega)$.

\textit{Application to FeSe.} To illustrate the utility of this new method, we consider the general question of impurity states in Fe-based superconductors, which often
exhibit unusual spatial forms.  In particular, geometric dimers, high intensity conductance or topographic features localized on the sites of the
two pnictide or chalcogenide atoms neighboring an Fe site, are ubiquitous in a number of Fe-based materials.
  It is not known whether these defects are Fe vacancies, adatoms or site switching, but they seem to be centered on
  a given Fe(I) or Fe(II) site, as imaged by STM topography, and it is natural to assume that the impurity state is
  coupled to the pnictide or chalcogenide atoms closest to the surface, i.e. the defects are oriented one way or
  another according to whether they are centered on Fe(I) or Fe(II).     We consider here the effect of a  repulsive potential
  on the Fe site, to crudely model a vacancy, site switching, or adatom, in the FeSe material ($T_c=8\,\text{K}$).   

\begin{figure}
\includegraphics[width=1\columnwidth]{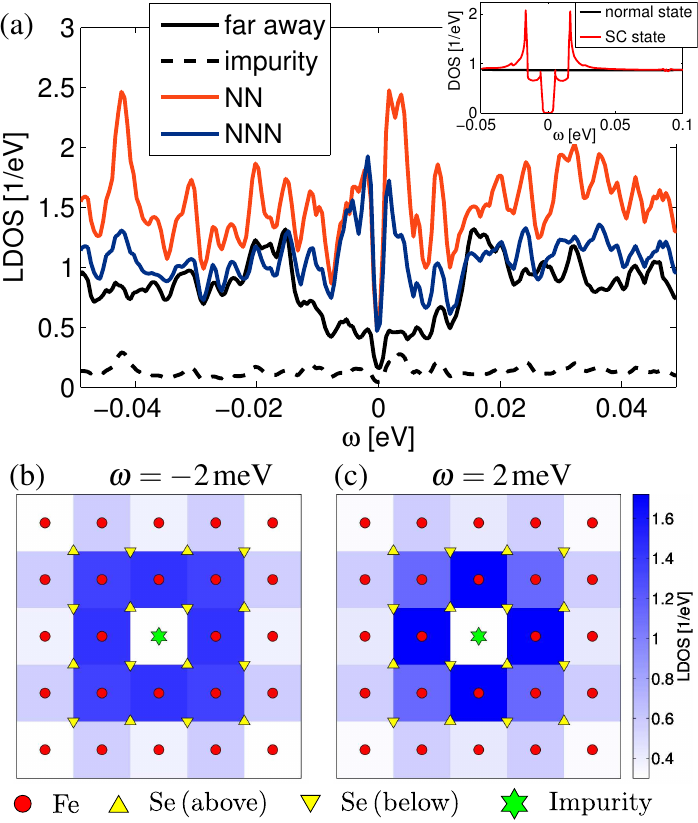}
\caption{(color online)  (a) density of states in SC state (inset: without impurity), far from
impurity (black), at impurity site (black, dashed), on nearest neighbor
site (orange [light gray]) and on next-nearest neighbor site (blue [dark gray]), calculated with a tetrahedron method using $40\times40$ supercells.
(b) resonant state real space BdG patterns at $\omega=- 2\,\text{meV}$
and (c) $\omega=2\,\text{meV}$. }
\label{fig:dos}
\end{figure}

      We now consider the form of the superconducting gap, which is not well-established for this system.  The STS tunneling for thin films of FeSe on graphite
  is V-shaped at low energies, implying the presence of gap nodes or small minimum gap.  We note, however, that Ref. \onlinecite{DLFeng}  has proposed that
  nodes may arise from a weak SDW state present in these  films.   This scenario is indeed consistent
  with the $\sim 16 a_0$ long ``electronic dimers'', high intensity spots observed with axes aligned at 45 degrees with respect to the geometrical dimers, which have been interpreted as emergent defect states in an ordered magnetic phase\cite{Navarro_cigar}.   However since we are primarily interested in exhibiting the local $C_4$ symmetry breaking effects due to ligand atoms possible within our current method, we ignore the possible
  effects of magnetism in this work, and calculate the real space pair potentials $\Gamma_{\mathbf{RR}'}^{\mu\nu}$  within the  random phase approximation (RPA) using the 10-orbital tight-binding band structure with  $U=0.90\,\text{eV}$ and $J=U/4$,~\cite{footnote} following the procedure described in Ref. \onlinecite{Navarro_LiFeAs}.   The 10-orbital BdG equations are then solved on $15\times15$ unit cell lattices with stable solutions found through iterations of the  self-consistency equation for the real space gaps  $\Delta_{\mathbf{RR'}}^{\mu\nu}=\Gamma_{\mathbf{RR}'}^{\mu\nu}\sum_{n}u_{\mathbf{R}\mu}^{n}v_{\mathbf{R'}\nu}^{n*}f(E_{n})$,  where $\sum_n$ denotes summation over all eigenstates $n$. When calculating the Green's function we use $20\times20$ supercells to acquire spectral resolution of order $\sim 0.5\,\text{meV}$.
The superconducting ground state is found to be of the usual multiband $s_\pm$ type.     Its DOS, shown as the red curve in the inset of Fig. \ref{fig:dos}(a),
reflects the several different orbital gaps in the problem, and  bears a striking resemblance to the nearly
V-shaped conductance observed in experiment\cite{song11}.  Note, however, that the overall gap magnitude is
much larger than in experiment, since we have chosen artificially large interaction parameters to deliberately create a unrealistically  large gap, so that  impurity bound states may be more clearly visualized within our numerical resolution.  Since the normal  state density of states for this material is flat at low energies, we do not expect this
to alter our results qualitatively, although the exact value of the impurity potential which creates a bound state
for a realistic gap size will change.   Finally, we observe that the choice of method used to calculate the gap does not influence our qualitative results, e.g. a single constant next-nearest neighbor pair potential that gives an $s_\pm$ state is sufficient (see Supplement Material).

\textit{Single impurity in FeSe.} We now add a single impurity to the system, of strength  $V_\text{imp}=5\,\text{eV}$  in all orbital channels $\mu$,
which creates an impurity bound state within the spectral gap at $\pm \Omega_0\simeq 2\,\text{meV}$ (Fig. \ref{fig:dos}(b)).
 The value chosen for  this potential is not based on a microscopic description of a particular defect, but merely to create an in-gap bound state, to illustrate the method.
Such bound states are not universal, but depend on the electronic structure
of the host, impurity potential details, and gap functions,
as has been emphasized in Refs.~\onlinecite{HKM_review,Vekhter1imp}.  The structure of this complex bound state extends
above and below the Fe plane, as can now be visualized by the current method.

 \begin{figure}[t]
\includegraphics[width=1\columnwidth]{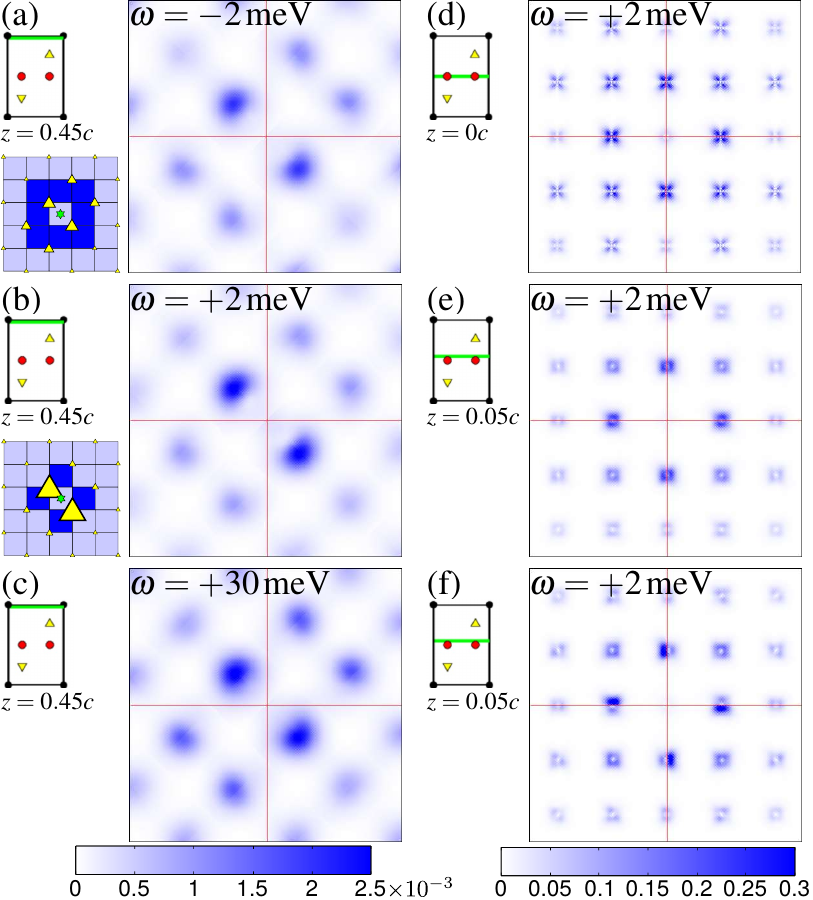}
\caption{ (color online) xy cuts through continuous 3D LDOS$(x,y,z;\omega)$ in $(\text{eV}\,\text{bohr}^3)^{-1}$ at
	at the surface, $z=0.45c$ and different energies (a) $\omega=-2\,\text{meV}$, (b) $2\,\text{meV}$, (c) $30\,\text{meV}$, and heights at (close to) the Fe-plane $z=0c$ (d), $z=0.05c$ (e,f) and the same energy $2\,\text{meV}$, where the $c$-axis lattice constant is $c=10.44\,\text{bohr}$.  All maps are calculated from Eq. (\ref{eq:continuumG}) except for
 (e), which includes only the local ($\R=\R'$, $\mu=\nu$) contributions.
The schematic side views of the unit cell indicate the $z$-value of the cut (green line)
relative to
the Fe (red circles) and Se (yellow triangle) positions.
The  cartoon  left of  (a,b) explains the LDOS patterns in terms of the resonant nearest neighbor Fe Wannier orbitals (compare Fig. \ref{fig:dos}(b,c)) and their upper Se tails.  Thin red lines are directed along Fe-Fe bonds through the central impurity site, and the black border indicates the extent of the 5$\times$5 Fe region.
}
\label{fig:ldos}
\end{figure}

 In Fig. \ref{fig:dos} (b), (c),  we show first the local lattice density of states $\rho_\text{BdG}(\R,\omega)\equiv -\frac 1 \pi \mathop\text{Im}\, G(\R,\R;\omega)$ obtained at the positive and negative resonance
 energies within a conventional 10-Fe orbital BdG calculation.  As is clear, both resonance patterns are $C_4$ symmetric
 as they must be for the tetragonal FeSe system, and do not resemble any defect states imaged in STS experiments on this system.
 The orthorhombic distortion of FeSe at low temperatures has been disregarded in our work, as it cannot account for the $C_4$ to $C_2$ symmetry breaking close to the impurity with axis 45 degrees away from the Fe-Fe bond.
 In Fig.  \ref{fig:ldos}, we finally plot the analogous continuum LDOS $\rho(\rr,\omega)\equiv -\frac 1 \pi \mathop\text{Im}\, G(\rr,\rr;\omega)$, which extends in three dimensions, although the set of $\R$ considered in Eq. (\ref{eq:continuumG}) lies in a 2D plane, due to the 3D spatial extent of the Wannier functions. Cuts at various heights $z$ from $z=0$ in the Fe-Fe plane to a point very close to the Se plane are shown.  As expected from Fig.~\ref{fig:dos} (b), (c), in the Fe
   plane $\rho(\rr,\omega)$ is still $C_4$ symmetric (Fig.~\ref{fig:ldos} (d)), but when one increases $z$ the local placement of the Se atoms
   breaks this symmetry to $C_2$, as clearly seen in  Fig.~\ref{fig:ldos} (a-c) for $z$ close to the surface  where the STM tip is roughly located.
 In our result, the dimer from the large LDOS close to the NN (up) Se atoms is visible at all energies, but 
 there is some signicant variation with energy in the intensity on the NNN (up) Se, yet this may change according to the details of the actual defect potential.
Looking at the corresponding Fe-only BdG LDOS patterns in Fig. \ref{fig:dos} (b-c), also shown as cartoon in Fig.~\ref{fig:ldos}  (a-b),
 suggests a simple explanation of the patterns, namely that intensity maxima occur on those Se sites associated with the
 resonant Fe Wannier orbitals, with intensities on these Se sites adding  constructively.
 Of course,
 deviations are in general to be expected, since simply adding intensities from resonant sites neglects  contributions from nonresonant ones,  as well as from nonlocal terms $\mathbf{R}\ne\mathbf{R'},\nu\ne\mu$ in the continuum Green's function arising in Eq. (\ref{eq:continuumG}).
 The importance of these  latter terms is illustrated  in Fig. \ref{fig:ldos}(e), where the
 local, orbitally diagonal contributions are plotted alone for a cut just above the Fe plane.
 The true, significant symmetry breaking is only recovered in the full result Fig.~\ref{fig:ldos}(f).

\begin{figure}
\includegraphics[width=1\columnwidth]{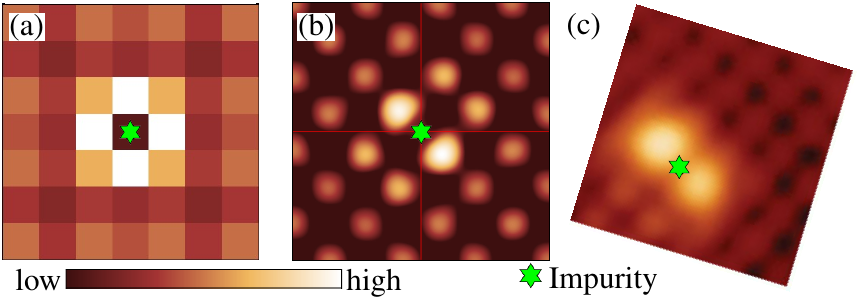}
\caption{(color online) Topography of impurity state: comparison of topograph with setpoint bias  $+6\,\text{mV}$
 for  (a) BdG calculation of impurity state as in Fig. \ref{fig:dos}(b); and (b) BdG-Wannier calculation of same impurity state.  In (c) we reproduce the experimental topograph from Ref. \onlinecite{song12},
rotated such that the directions of the lattice vectors match in all images.
 }
\label{fig:comparison}
\end{figure}

\textit{Discussion.}
To illustrate the dramatic improvements provided by our proposed approach we compare in Fig. \ref{fig:comparison} the topographs of the impurity state calculated within the BdG and BdG-Wannier approaches with the experimentally observed \cite{song12} geometrical dimer \cite{footnote2}.
The BdG-Wannier approach of course will also reproduce the experimentally observed \cite{song12} 90 degree rotation of the dimer when the impurity potential is moved from the Fe(I) sublattice to the Fe(II) sublattice.
Other comparisons with experiment are difficult, as detailed spectral features of these states are not yet published. 
The low-energy resonant state created by the opening of the superconducting gap (Fig. \ref{fig:ldos}(a,b,d,f)), and a nonresonant impurity state with similar patterns visible at higher energy (Fig. \ref{fig:ldos}(c)), broader
in energy due to its coupling to the metallic continuum, represent predictions which can be tested by experiment when spectral data are available.

 \textit{Conclusions.}  We have introduced a method to enable theoretical visualization of inhomogeneous states in superconductors,
 combining traditional solutions of the Bogoliubov-de Gennes equations with a first principles Wannier analysis.   The method
 not only enables a much higher spatial resolution, but also captures  the local symmetry internal to the unit cell of the crystal.
 Furthermore the method incorporates the nonlocal lattice Green's function contributions, which we have demonstrated to be of qualitative importance.
 As an example, we showed how ``geometric dimer" impurity states seen in Fe-based superconductors can be understood
 as  consequence of simple defects located on the Fe site due to the hybridization with the pnictogen/chalcogen states.
In terms of both symmetry and higher spatial resolution, the result obtained with the method introduced here represents a qualitative
improvement over conventional BdG investigations (Fig. \ref{fig:comparison}), and  opens a new window on the
theoretical analysis of  atomic scale phenomena in superconductors.

\textit{Acknowledgements.}  The authors are grateful to B.M. Andersen, H.P.-Cheng, M.N. Gastiasoro, J. Hoffman,  W. Ku, C.-L. Song, and  Y. Wang for useful discussions.
PC,  AK, and PJH were  supported by DOE DE-FG02-05ER46236, CC by NSFC 11274006, and TB  by DOE CMCSN DE-AC02-98CH10886 and as a Wigner Fellow at the Oak Ridge National Laboratory.  The authors are grateful to W. Ku for use of his Wannier function code.

\end{document}


\title{[Supplement Material]\\
Visualization of atomic-scale phenomena in superconductors: application to FeSe }
\author{Peayush Choubey$^1$, T. Berlijn$^{1,2}$, A. Kreisel$^1$, C. Cao$^3$ and  P. J. Hirschfeld$^1$}
\affiliation{
$^1$Department of Physics, University of Florida, Gainesville, Florida 32611, USA\\
$^2$Center for Nanophase Materials Sciences and Computer Science and Mathematics Division, Oak Ridge National Laboratory, Oak Ridge, Tennessee 37831, USA
\\
$^3$Condensed Matter Physics Group, Department of Physics,
Hangzhou Normal University, Hangzhou 310036, China}

\date{\today}

\maketitle

In this supplementary material we give some details on our application of the method to the FeSe system.\\
 \renewcommand{\thefigure}{\Roman{figure}}
\textit{Density Functional calculations.}
As mentioned in the main text, we use Density Functional Theory (DFT) to obtain a realistic model for our
calculations using the Bogoliubov-de Gennes method as well as the construction of the retarded continuum Green' function.
The starting point is the experimental crystal structure of bulk FeSe as reported in Ref.~\cite{Margadonna08} with
symmetry group $ P4/nmm$ and lattice constants $a=b=7.13\,\text{bohr}$, $c=10.44\,\text{bohr}$ as well as the fractional position $z= 0.265$ of the Se atoms.
\begin{figure}[tb]
\includegraphics[width=1\columnwidth]{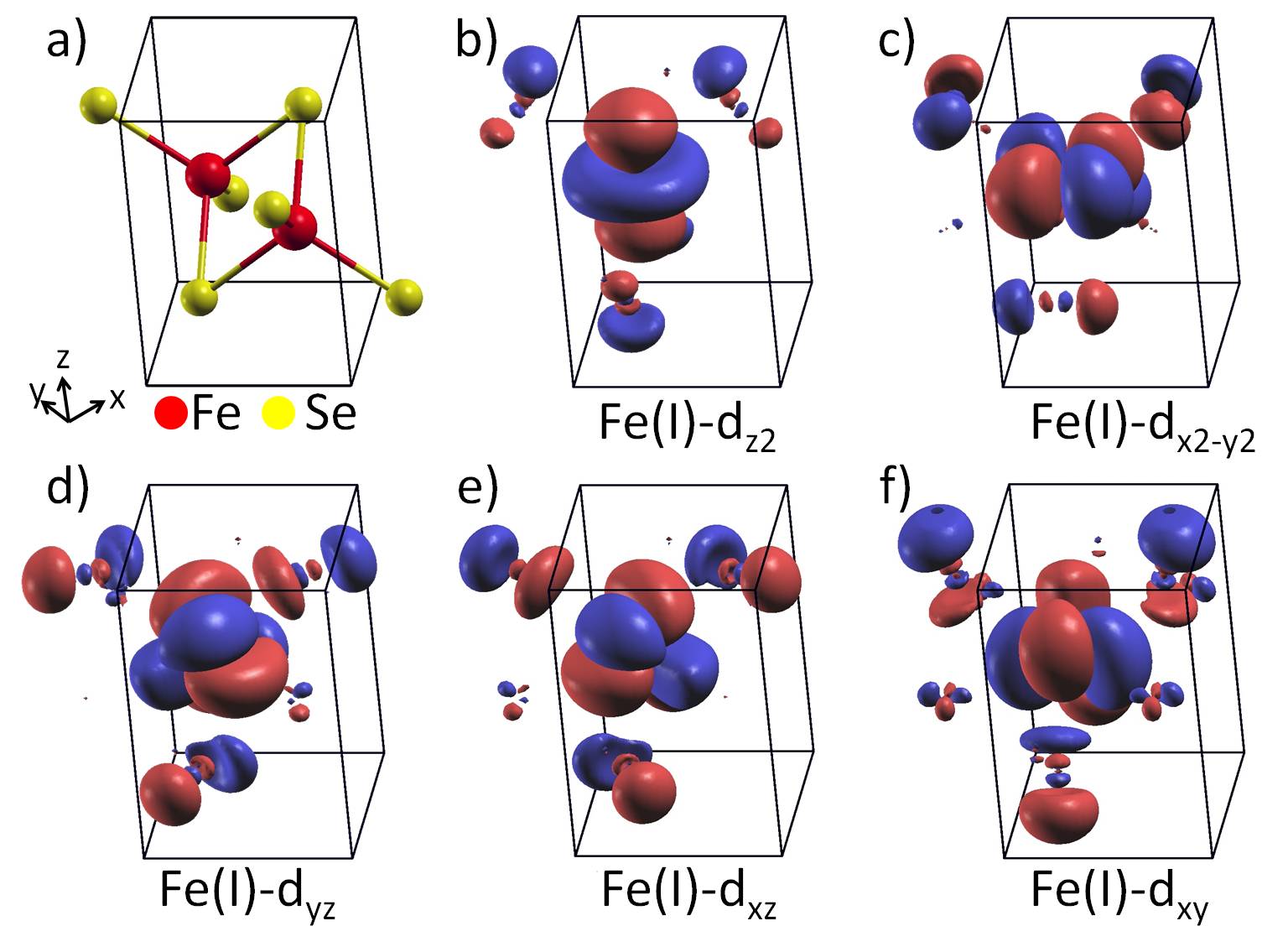}
\caption{(Color online)  Isosurface plots of Fe-$d$ Wannier orbitals in FeSe at $0.03\,\text{bohr}^{-3/2}$;  red and blue indicate   phase of the wave function.  a) Side view of unit cell containing two Fe atoms (red) and two Se atoms (yellow), plus four Se    exterior to unit cell.   b) -f) Side views of $d$ Wannier orbitals on Fe(I) site.}
\label{fig_wannier}
\end{figure}
First principles calculations of the electronic structure using the Wien2K\cite{Blaha} package are followed
by a projected Wannier method preserving the local symmetry\cite{Ku_Wannier,Anisimov}.
 The Wannier orbitals with dominant $d_{xy}$ character are shown in Fig.~1 of the main text while a choice
of other Wannier orbitals that give smaller contributions to the local density of states at small energies
are visualized in Fig.~\ref{fig_wannier}.
Next, the tight binding Hamiltonian is restricted to a two-dimensional model Hamiltonian by ignoring the
hopping matrix elements in the z-direction while keeping the filling in the ten-orbital model fixed to $n=6.0$ which
corresponds to an average of the Hamiltonian matrix when partially Fourier transformed with respect to only the
$z$-coordinate. The Fermi surface of the corresponding model Hamiltonian and its orbital character are shown in Fig.~\ref{fig_fermi}.\\

\begin{figure}[tb]
\includegraphics[width=0.9\columnwidth]{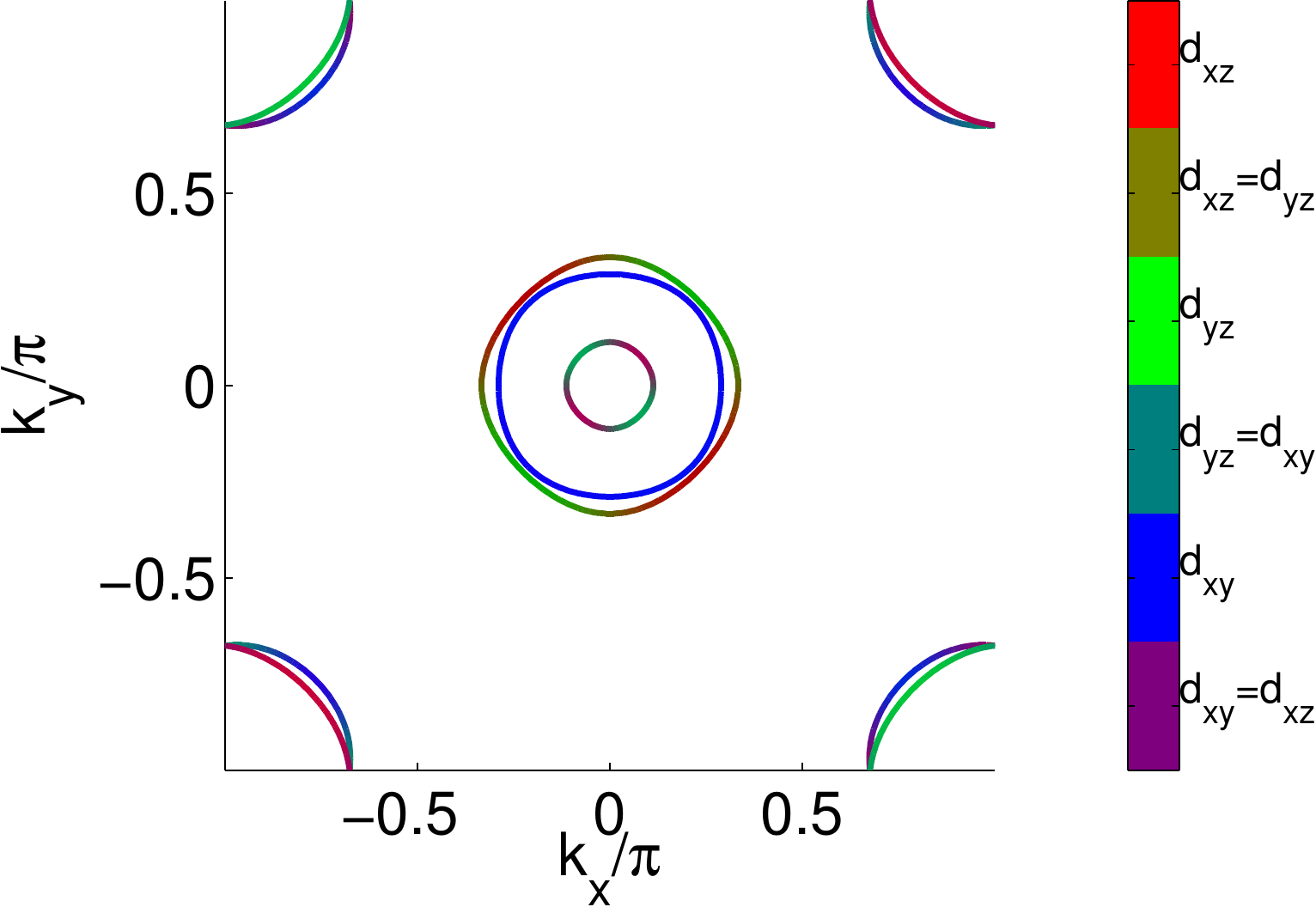}
\caption{(Color online) Orbital character on the Fermi surface of our two dimensional tight-binding Hamiltonian for FeSe, red $d_{xz}$, green $d_{yz}$, and blue $d_{xy}$ visualized with the summed-color method where the absolute value of the overlap is mapped to the RGB value of the color on the surface.}
\label{fig_fermi}
\end{figure}
\textit{Spin fluctuation theory.}
Next, we calculate the pairing interactions in real space starting from the tight-binding Hamiltonian
in conjunction with a Hubbard Hund Hamiltonian
\begin{align}
	H =&{U}\sum_{\mathbf{R},\mu}n_{\mathbf{R}\mu\uparrow}n_{\mathbf{R}\mu\downarrow}+{U}'\sum_{\mathbf{R},\nu'<\mu}^\prime n_{\mathbf{R}\mu}n_{\mathbf{R}\nu}
	\nonumber\\
	 + & {J}\sum^\prime_{\mathbf{R},\nu<\mu}\sum_{\sigma,\sigma'}c_{\mathbf{R}\mu\sigma}^{\dagger}c_{\mathbf{R}\nu\sigma'}^{\dagger}c_{\mathbf{R}\mu\sigma'}c_{\mathbf{R}\nu\sigma}\\
	 + & {J}'\sum^\prime_{\mathbf{R},\nu\neq\mu}c_{\mathbf{R}\mu\uparrow}^{\dagger}c_{\mathbf{R}\mu\downarrow}^{\dagger}c_{\mathbf{R}\nu\downarrow}c_{\mathbf{R}\nu\uparrow} \nonumber
\end{align}
\begin{figure}[tb]
\includegraphics[width=1\columnwidth]{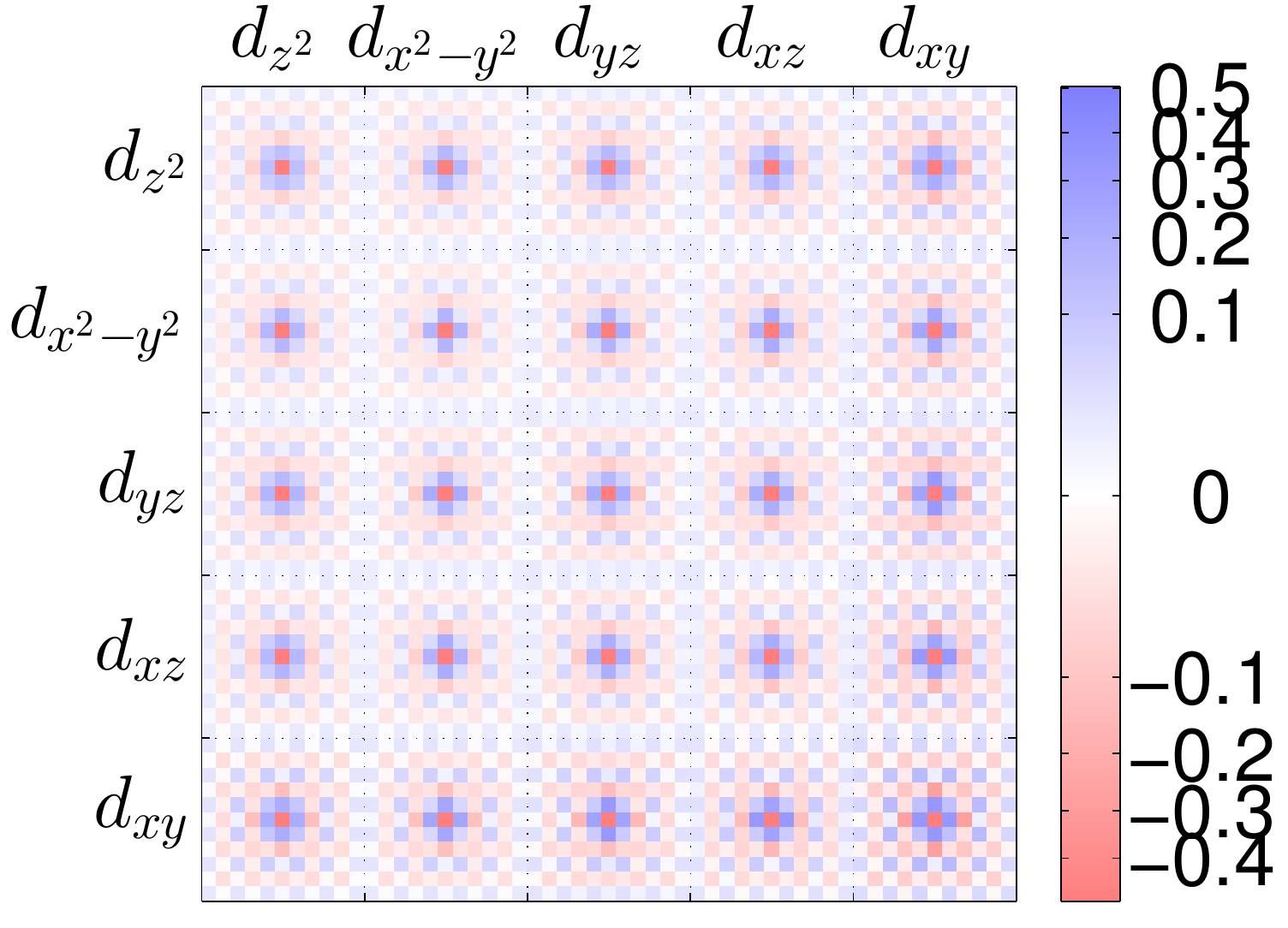}
\caption{(Color online) Pairing interaction in real and orbital space  $\Gamma_{\mathbf{RR'}}^{\mu\nu}$ (units in eV) calculated using the interaction parameters $U=0.90\,\text{eV}$ and $J=U/4$ and visualized using one pixel per Fe site. Note that the diagonal terms are larger and all terms are dominated by the on-site repulsion which varies from $1.2\,\text{eV}$ to $4\,\text{eV}$ and is cropped in the given scale.}
\label{fig_gamma}
\end{figure}
where the interaction parameters ${U}$, ${U}'=U-2J$, ${J}$, ${J}'=J$ are given in the notation of Kuroki \textit{et al.} \cite{Kuroki08} and the $\sum^\prime$ only gives a contribution when the indices $\mu$ and $\nu$ label an orbital on the same iron atom.
The pair scattering amplitudes in momentum space are calculated using the formula
\renewcommand{\ell}{\mu}
\begin{align}
	&{\Gamma}_{\ell_1\ell_2\ell_3\ell_4} (\k,\k') = \left[\frac{3}{2} \bar U^s \chi_1^\text{RPA} (\k-\k') \bar U^s\nonumber \right.\,~~~~~~\,\\
	&\,~~~\left. +  \frac{1}{2} \bar U^s - \frac{1}{2}\bar U^c \chi_0^\text{RPA} (\k-\k') \bar U^c + \frac{1}{2} \bar U^c \right]_{\ell_1\ell_2\ell_3\ell_4}. \label{eq:fullGamma}
\end{align}
where the charge and spin susceptibilities have been calculated within the random phase approximation (RPA)
\begin{subequations}
\begin{align}
\chi_{1\,\ell_1\ell_2\ell_3\ell_4}^\text{RPA} (\q) &= \left\{ \chi^0 (\q) \left[1 -\bar U^s \chi^0 (\q) \right]^{-1} \right\}_{\ell_1\ell_2\ell_3\ell_4},\\
 \chi_{0\,\ell_1\ell_2\ell_3\ell_4}^\text{RPA} (\q) &= \left\{ \chi^0 (\q) \left[1 +\bar U^c \chi^0 (\q) \right]^{-1} \right\}_{\ell_1\ell_2\ell_3\ell_4}.\label{eqn:RPA}
\end{align}
\end{subequations}
from the bare susceptibilities $\chi^0 (\q)$ and the definition of the interaction matrices $\bar U^s$ and $\bar U^c$ can be found in Ref.~\cite{a_kemper_10}. The dominant pairing interaction in orbital and real space is then obtained by a Fourier transformation of the pair scattering amplitudes projected to the spin-singlet channel
\begin{equation}
 \Gamma_{\mathbf{RR'}}^{\mu\nu}=\frac 12 \sum_{\mathbf{k}} [ {\Gamma}_{\mu\nu\nu\mu} (\k,-\k)+{\Gamma}_{\mu\nu\nu\mu} (\k,\k)] e^{-i\k\cdot(\mathbf{R}-\mathbf{R}')}\,,
\end{equation}
where the small matrix elements ${\Gamma}_{\mu\nu\mu\nu} (\k,-\k')$ have been neglected for simplicity\cite{a_kemper_10}. As shown in Fig. \ref{fig_gamma}, the pairing interaction shows a rapidly decaying checkerboard pattern as a function of spatial distance such that we safely can truncate it with few unit cells.\\

\begin{figure}[tb]
\includegraphics[width=1\linewidth]{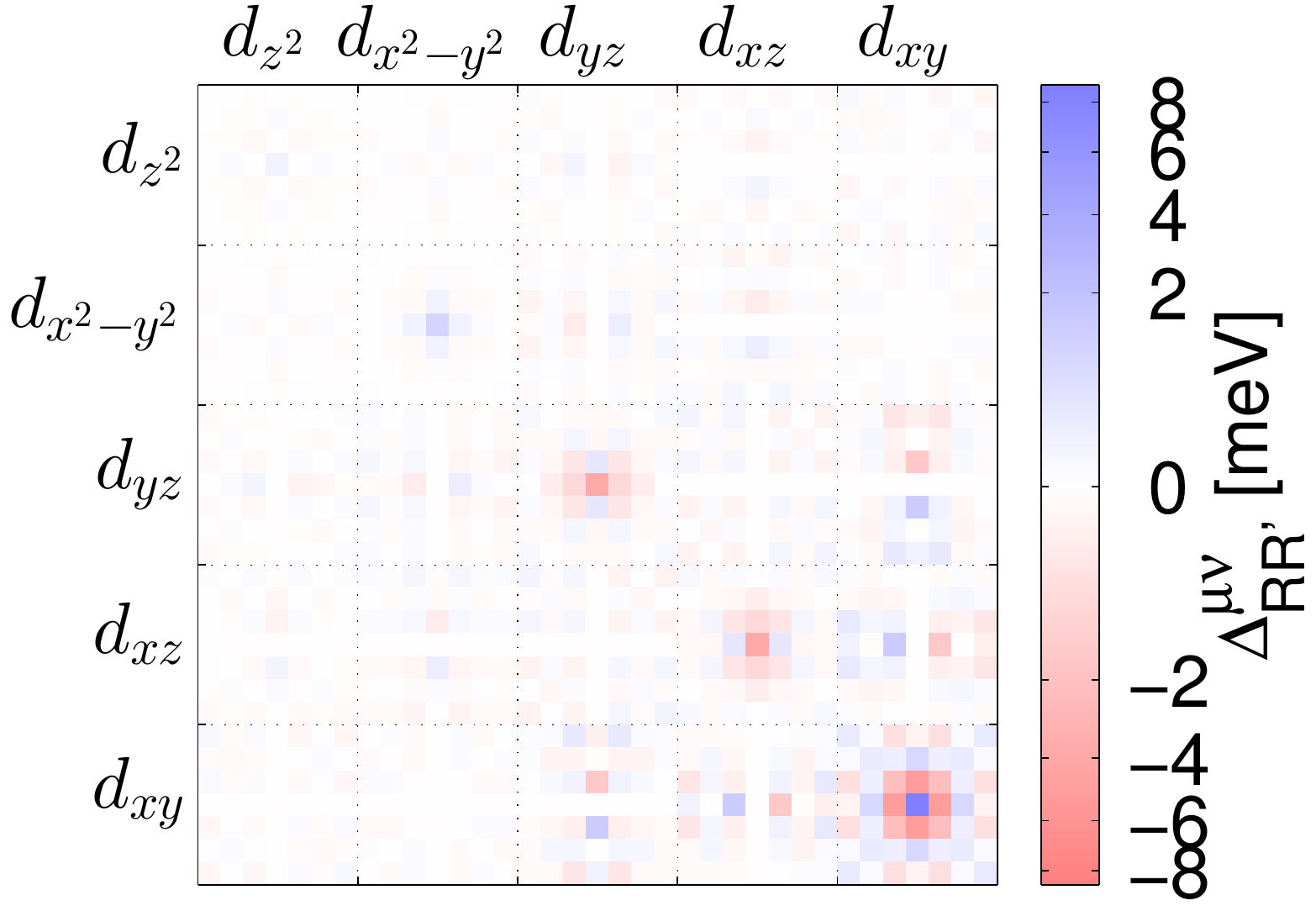}
    \rput[tr](-0.28\linewidth,0.7\linewidth){a)}
\newline
\includegraphics[width=0.9\linewidth]{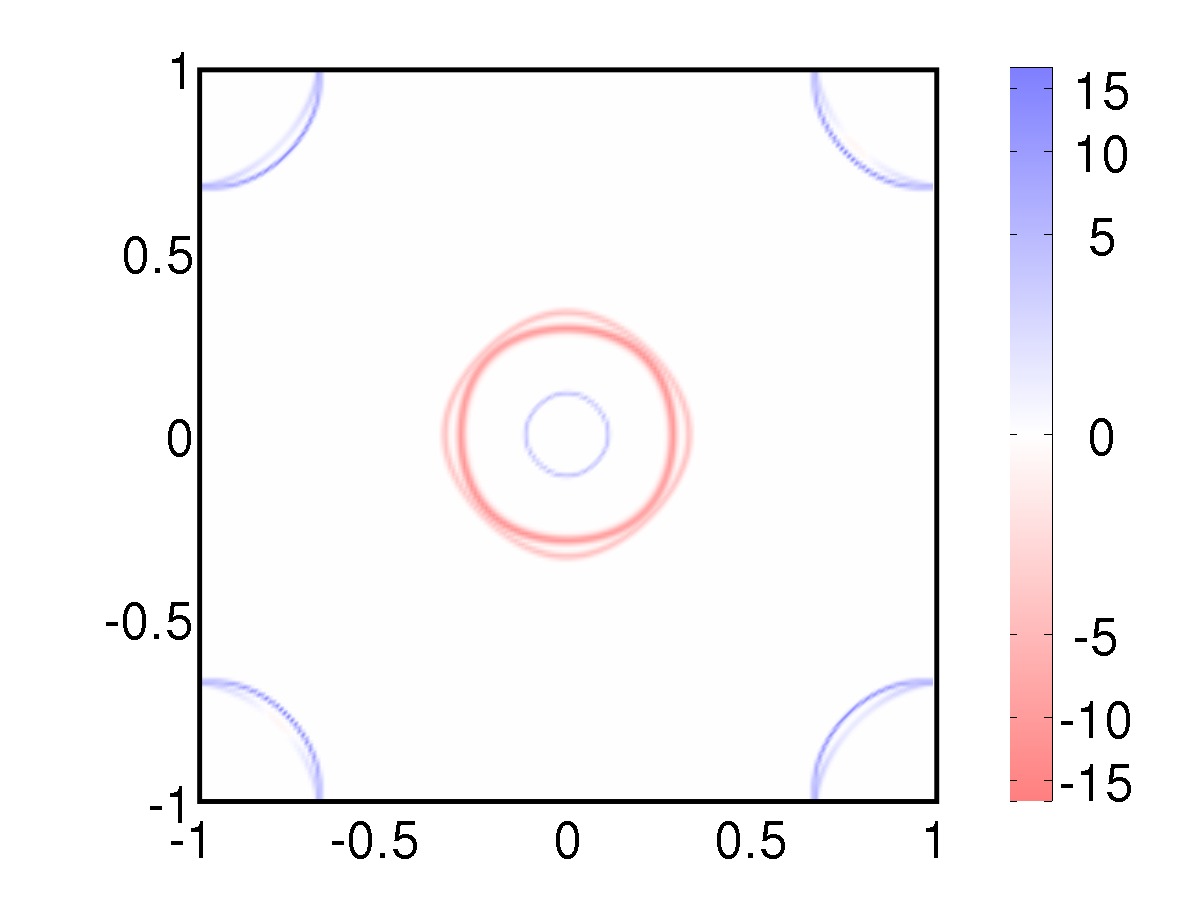}
    \rput[tr](-0.88\linewidth,0.7\linewidth){b)}
    \rput[tr](-0.83\linewidth,0.38\linewidth){$ {k_x}/\pi$}
    \rput[tr](-0.45\linewidth,0.01\linewidth){$ {k_y}/\pi$}
    \rput[tr](-0.475\linewidth,0.37\linewidth){$\Gamma$}
    \rput[tr](-0.18\linewidth,0.66\linewidth){$M$}
    \rput[tr](0.0\linewidth,0.44\linewidth){\rotatebox{90}{$\Delta(\k)\,[\text{meV}]$}}
\caption{(Color online) a) Superconducting gap $\Delta_{{\mathbf{RR'}}}^{\mu\nu}$ in meV as obtained from the self-consistent solution of the Bogoliubov-de Gennes equation and b) the corresponding gap $\Delta(\k)$ in momentum space projected on the Fermi surface with a Gaussian energy broadening of $8\,\text{meV}$. Note that the scale is nonlinear to make features of the gap visible that are small in magnitude.}
\label{fig_gap}
\end{figure}

\textit{Superconducting gap.}
The interaction parameters \mbox{$U=0.90\,\text{eV}$} and $J=U/4$ are now chosen such that the Bogoliubov-de Gennes Hamiltonian as given in the main text shows a robust instability in the superconducting channel. The real space pattern of the gap structure $\Delta_{{\mathbf{RR'}}}^{\mu\nu}$ can be considered as independent of the system size starting with $N=15$ unit cells.
The gap structure in orbital space reflects the symmetries of the underlying Fe $d$ states and is rather short range as shown in Fig.~\ref{fig_gap} a). The Fourier transform of the orbital gap transformed to band space using the normal-state eigenvectors of the bare Hamiltonian $H_0$ displayed Fig.~\ref{fig_gap} b) reflects the small gap features on the $\Gamma$ centered inner pocket while the gaps on the middle $\Gamma$ centered pocket and the large $M$-centered pocket are rather large. Note that the gap on the inner hole pocket shows a sign change with respect to the other hole pockets around $\Gamma$. \\

\def \rputx {-0.45}
\def \rputy {0.69}
\textit{Calculations using phenomenological interactions.}
 \begin{figure}[tb]
\includegraphics[width=1\columnwidth]{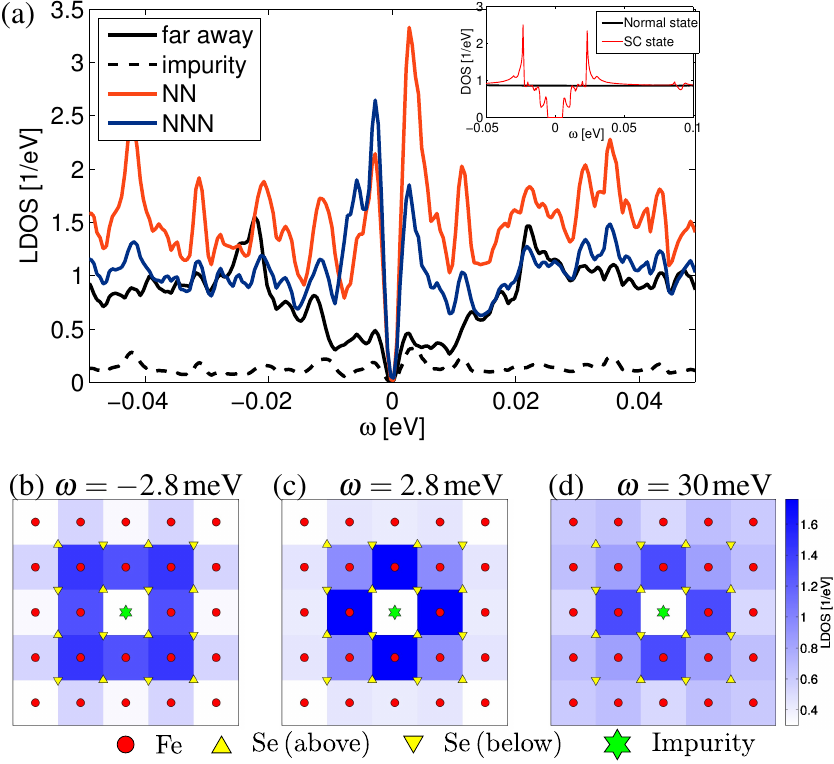}
\caption{(color online)  a) density of states in SC state (inset: without impurity), far from
impurity (black), at impurity site (black, dashed), on nearest neighbor
site (orange [light gray]) and on next-nearest neighbor site (blue [dark gray]), calculated with a tetrahedron method using $40\times40$ supercells for the phenomenological NNN pairing interaction.
(b) resonant state real space BdG patterns at $\omega= -2.8\,\text{meV}$, (c) $\omega= 2.8\,\text{meV}$, 
and (d) $\omega=30\,\text{meV}$. }
\label{fig:dos}
\end{figure}
In order to show that the results of our method do not strongly depend on the microscopic details of the underlying model describing superconductivity, we additionally performed similar calculations with phenomenological pairing interactions instead of the interactions generated from the fluctuation exchange mechanism. To achieve a similar sign-changing s-wave gap structure we used only interactions to the second nearest neighbors (NNN) such that the nonzero pair scattering amplitudes read
\begin{equation}
 \Gamma_{\mathbf{RR'}}^{\mu\nu}=\Gamma_0 \delta_{\mu,\nu}\label{eq:pairing}
\end{equation}
for $\mathbf{R}-\mathbf{R'}$ being a vector connecting NNN iron sites, with a constant $\Gamma_0=0.3\,\text{eV}$. Solving the BdG equation for a system with $N=15$ elementary cells yields a gap structure in real space with NNN gaps  of the order of $6\,\text{meV}$ on the $d_{xy}$ orbital and the components of the other orbitals  smaller as also observed for the calculation with spin fluctuation mediated pair interactions. As seen in Fig. \ref{fig:dos}, density of states in the superconducting state is nearly identical to the one shown in the main text (Fig. 2 a) inset). The same also applies for the local density of states in presence of an impurity with potential $V_{\text{imp}}=5\,\text{eV}$, except that the peak of the impurity bound state is  more pronounced and shifted to a slightly larger absolute energy $\pm \Omega_0=2.8\,\text{meV}$.
 Note also that the real space BdG patterns are basically the same as the corresponding result shown in the main text, anticipating that the actual result is not very sensitive to the choice of the pairing interaction as long as the resulting superconducting gap in the homogeneous case is similar.
\begin{figure}[tb]
\includegraphics[width=1\columnwidth]{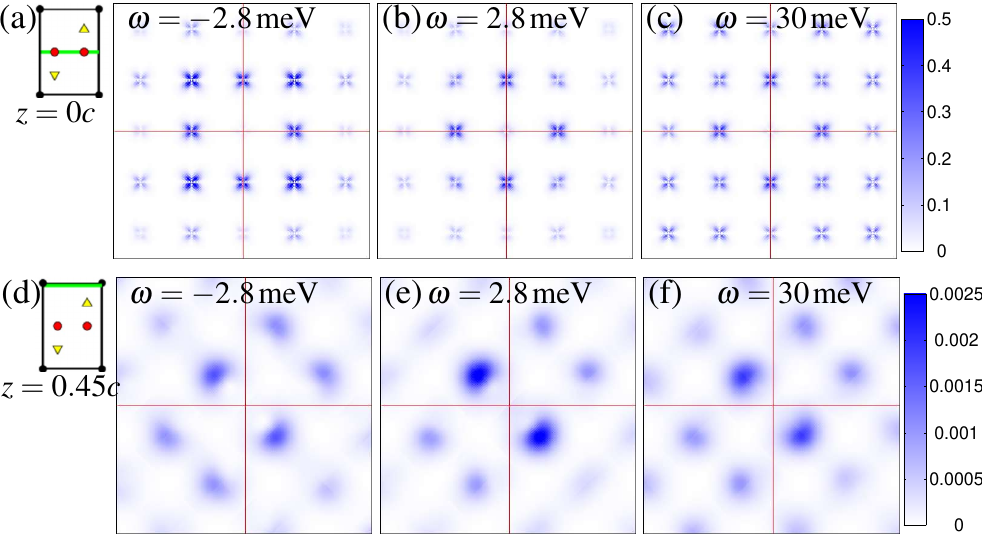}
\caption{ (color online) xy cuts through continuous 3D LDOS$(x,y,z;\omega)$
in $(\text{eV}\,\text{bohr}^3)^{-1}$
at  $\omega=-2.8\,\text{meV}$ (a,d),$\omega=2.8\,\text{meV}$ (b,e), $30\,\text{meV}$ (c,f) and $z=0c$ (first row), $z=0.45c$ (second row), where $c$ is the $c$-axis lattice constant $10.44\,\text{bohr}$.  All maps are calculated using the phenomenological pairing parameters given in Eq.  (\ref{eq:pairing}).
The schematic side views of the unit cell indicate the z-value of the cut (green line)
relative to the Fe (red circles) and Se (yellow triangle) positions.
Thin red lines are directed along Fe-Fe bonds through the central impurity site, and the black border indicates the extent of the 5$\times$5 Fe region.}
\label{fig:ldos}
\end{figure}
To 
demonstrate further that the results of our BdG-Wannier method are robust against choices of model parameters, we  recalculated the maps of the continuous 3D LDOS at the same cuts in z-direction as in the main manuscript. Fig. \ref{fig:ldos} presents the corresponding maps obtained from the calculation using the phenomenological pairing parameters and is very similar to the result shown in the main text: At the Fe-plane ($z=0c$),
one sees primarily the Fe-d orbitals, however at the relevant height ($z=0.45c$), the resonant behavior of the Se tails of the Wannier functions are clearly visible as a geometric dimer.\\

\textit{Calculation of the topography.}
In STM experiments at least two different methods can be used to image the real space properties of a superconductor:   topography and  conductance maps.  We have discussed the calculation of conductance maps in the main text.
In order to obtain  topographic map using our method, we started from the approximate formula for the tunneling current $I$ at a given bias voltage $V$\cite{Hoffman2011}
\begin{equation}
I(V,x,y,z)=-\frac{4\pi e}{\hbar} \rho_t(0) |M|^2 \int_0^{eV} \rho(x,y,z,\epsilon) d\epsilon\,,
\end{equation}
where $x,y,z$ are the coordinates of the tip, $\rho(x,y,z,\epsilon)$ is the continuum LDOS as defined in the main text, $\rho_t(0)$ is the density of states of the tip, and $|M|^2$ is the square of the the matrix element for the tunneling barrier. For the topography, we now have to impose the condition $I(V,x,y,z)=\text{const.}$ which means to find the contour of constant integrated local density of states. We therefore calculated all LDOS maps $\rho(x,y,z,\epsilon)$ for $z=0\ldots0.7 c$ and energies $\epsilon=0\ldots 6\,\text{meV}$, performed the integration over energy and calculated the height $z(x,y)$ where
$\int_0^{eV} \rho(x,y,z,\epsilon) d\epsilon=3.45 \cdot 10^{-7}$. Fig. 4 b) of the main text shows the result $z(x,y)$ in a color-scale plot with a color scheme which approximately matches the one from the experimental result\cite{song11}.